\documentstyle[12pt,aaspp4]{article}

\def\cosmo{H$_0$ = 50~km~s$^{-1}$~Mpc$^{-1}$ and q$_0 = 0$ } 
\def\en{$\;$}

\received{August 11, 1999}
\accepted{}
\journalid{}{}
\articleid{}{}

\begin{document}
\input{psfig.tex}

\title{\bf The HST Survey of BL~Lacertae Objects. II. Host Galaxies}

\author{C. Megan Urry, Riccardo Scarpa,
	and Matthew O'Dowd\altaffilmark{1}}
\authoraddr{Space Telescope Science Institute, 3700 San Martin Dr., 
Baltimore, MD 21218, USA}
\authoremail{cmu@stsci.edu,scarpa@stsci.edu,odowd@stsci.edu}
\affil{Space Telescope Science Institute}
\altaffiltext{1}{Also at School of Physics, University of Melbourne, 
Parkville, Victoria, Australia 3052}

\author{Renato Falomo} 
\authoraddr{Osservatorio Astronomico di Padova, Vicolo dell'Osservatorio 5, 
35122 Padova, Italy}
\authoremail{falomo@astrpd.pd.astro.it}
\affil{Astronomical Observatory of Padova}

\author{Joseph E. Pesce\altaffilmark{2}} 
\authoraddr{Department of Astronomy, Pennsylvania State University,
525 Davey Lab, University Park, PA 16802 }
\authoremail{pesce@astro.psu.edu}
\affil{Pennsylvania State University}
\altaffiltext{2}{Also, Eureka Scientific}

\author{Aldo Treves} 
\authoraddr{Universita' dell'Insubria, via Lucini 3, 22100 Como }
\authoremail{treves@uni.mi.astro.it}
\affil{University of Insubria}

\begin{abstract}

We have used the {\it HST} WFPC2 camera to survey 132 BL~Lac objects
comprising seven complete radio-, X-ray-, and optically-selected
samples. We obtained useful images for 110 targets spanning the
redshift range $0\lesssim z \lesssim 1.3$. These represent an unbiased subsample 
of
the original 132 since they were snapshots selected to fill random holes in the
{\it HST} schedule. The exposure times ranged from a few hundred to
$\sim1000$ seconds, increasing with redshift. Most images were taken
in the F702W filter; those already observed in F814W during Cycle 5
were re-observed in F606W to give broader wavelength coverage. The
data were analyzed uniformly, and both statistical and systematic
errors were estimated (the latter dominate).
In two thirds of the BL~Lac images, host galaxies are detected,
including nearly all for $z<0.5$ (58 of 63). 
In contrast, only one quarter of the BL~Lacs with $z>0.5$ 
(6 of 22)
were resolved because of the relatively
short exposure times, and these tend to be very luminous host
galaxies. The highest redshift host galaxy detected is in a BL~Lac
object at $z=0.664$.
{\it HST} data add critical morphological information in the range a few
tenths to a few arcseconds.
In 58 of the 72 resolved host galaxies, 
a de~Vaucouleurs profile is significantly preferred, at $\gtrsim 99$\% 
confidence,
over a pure exponential disk;
the two fits are comparable in the remaining 14 cases because
of their generally lower signal-to-noise ratios.
These results limit the number of disk systems to at most 8\% of BL~Lacs
(at 99\% confidence),
and are consistent with all BL~Lac host galaxies being ellipticals.

The detected host galaxies are luminous ellipticals with a median absolute
K-corrected magnitude of $M_R \sim -23.7 \pm 0.6 $~mag (rms 
dispersion), at least one magnitude brighter than $M^*$ and 
comparable to brightest cluster galaxies.
The galaxy morphologies are generally smooth and undisturbed,
with small or negligible ellipticities ($\epsilon\lesssim 0.2$). 
The half-light surface brightness is anti-correlated with
half-light radius in quantitatively the same way as other elliptical galaxies,
indicating that apart from their highly active nuclei, BL~Lacs appear to be
absolutely normal ellipticals.
There is no correlation between host galaxy and observed nuclear magnitude or 
estimated jet power corrected for beaming. If black hole mass is correlated 
linearly with bulge mass in general, this implies a large range 
in Eddington ratio. 
The host galaxies of the radio-selected and X-ray-selected BL~Lacs
are comparable in both morphology and luminosity,
strongly suggesting that nuclear properties 
do not have a dramatic effect on large-scale host galaxy 
properties, or vice-versa.
BL~Lac objects have extended radio powers and host galaxy magnitudes very
much like those of FR~I galaxies, and quite distinct from FR~IIs, which instead
are more similar to quasars.
Thus the present data strongly support the unification picture
with FR~I galaxies constituting the bulk of the parent population of
BL~Lac objects.

{\underline{\em Subject Headings:}}
BL~Lacertae objects --- galaxies: structure --- galaxies: elliptical

\end{abstract}
\setcounter{footnote}{0}

\section{Introduction}
\label{sec:intro}

The {\it Hubble Space Telescope (HST)} has been used extensively to
study the host galaxies of AGN, primarily quasars and radio galaxies
with relatively high nuclear luminosities 
(Disney et al. 1995; McLeod \& Rieke 1995; Bahcall et al. 1997; 
Best, Longair \& R\"ottgering 1997; Hooper, Impey \& Foltz 1997; 
McCarthy et al. 1997; Ridgway \& Stockton 1997; 
Serjeant, Rawlings \& Lacy 1997; Boyce, Disney \& Bleaken 1999; 
McLeod, Rieke \& Storrie-Lombardi 1999; McLure et al. 1999). 
Its order-of-magnitude better spatial resolution (over
a large field compared to adaptive optics) provides unique and
critical information at sub-arcsecond scales. 

Collectively, {\it HST}
and ground-based observations of host galaxies have already led to
interesting results. The idea that radio-loudness is uniquely related
to host galaxy type has been shown to be incorrect:
while radio-loud AGN are almost always found in elliptical
galaxies --- often luminous ellipticals comparable to brightest
cluster galaxies (Smith \& Heckman 1986; Hutchings, Janson \& Neff 1989; 
Veron-Cetty \& Woltjer 1990; Taylor et al. 1996; Zirm et al. 1998) ---
radio-quiet AGN are found in both elliptical or disk galaxies (Taylor
et al. 1996, Bahcall et al. 1997, McLure et at. 1999).
Several studies have reported that
the host galaxies of radio-quiet AGN are systematically less luminous
than those of radio-loud AGN (Smith \& Heckman 1986; 
Hutchings et al. 1989; Veron-Cetty \& Woltjer 1990; 
Lowenthal et al. 1995)
although this effect was not found in other samples matched for
redshift and luminosity (Taylor et al. 1996; Hooper et al. 1997). 
It has also been suggested that radio-quiet AGN
have less disturbed morphologies (Hutchings et al. 1989),
and certainly dust, tidal tails, and/or close companion galaxies are 
prevalent in radio-loud AGN (Smith \& Heckman 1986; Yee \& Green 1987; 
Bahcall et al. 1997; Canalizo \& Stockton 1997; Martel et al. 1997; 
Pentericci et al. 1999).

An interesting but controversial issue is the possible relation
between host galaxy magnitude and nuclear brightness. 
A trend for the brightest nuclei to lie in the most luminous galaxies
has been found in some 
(McLeod \& Rieke 1994, 1995; Hooper et al. 1997; McLeod et al. 1999) 
though not all
(Taylor et al. 1996; Wurtz, Stocke \& Yee 1996, hereafter WSY) 
host galaxy studies.
Such a correlation would indicate a close connection between
small-scale, black-hole-related phenomena and large-scale galactic
phenomena, possibly related to galaxy formation scenarios (e.g., Small
\& Blandford 1992; Haehnelt \& Rees 1993). Further, where black hole
masses have been reliably estimated in local galaxies they appear to
be proportional to the bulge mass (Kormendy \& Richstone 1995, 
Magorrian et al. 1998, van der Marel 1999);
if the efficiency of converting accreting mass into AGN
luminosity does not vary widely, then for AGN this would translate to
AGN luminosity being proportional to host galaxy magnitude (e.g.,
McLure et al. 1999).

A related issue is whether the cosmic evolution of galaxies and AGN 
is inextricably linked or essentially independent.
That is, does the central black
hole grow more or less independently of the stellar mass, or is there
significant feedback between small-scale and large-scale systems?
Many have noted the similarity of the cosmic evolution of star-forming
galaxies and of AGN (e.g., Silk \& Rees 1998) --- both peak somewhere
in the redshift range $z=1-3$ --- possibly indicating a close
connection between galaxy and black hole evolution. So far, there is
evidence, at least in radio-loud AGN, that AGN host galaxies contain
old stars, as if the galaxy formed at high redshift (Dunlop et al. 1996; 
Ridgway \& Stockton 1997; Best et al. 1998; De Vries et al. 1998), 
close to the epoch of quasar dominance (Foltz et al. 1992; 
Maloney \& Petrosian 1999). It is not yet clear whether this is the case for all
radio-loud AGN or only for the most luminous (those with the most massive 
black holes).

Despite extensive observations, the results to date on host galaxies
are surprisingly mixed, possibly because the samples studied contain
mostly higher luminosity AGN and are often far from complete, in part
because of selection effects. We therefore
undertook an {\it HST} snapshot survey\footnote{Based on observations
made with the NASA/ESA Hubble Space Telescope, obtained at the Space
Telescope Science Institute, which is operated by the Association of
Universities for Research in Astronomy, Inc., under NASA contract
NAS~5-26555.} to investigate the morphology, color, and luminosity of
AGN host galaxies, and the evolution of these properties with cosmic
epoch. Our strategy was to complement existing investigations by
investigating lower luminosity AGN out to moderate redshifts, $z\sim 1$.
We also concentrated on radio-loud AGN, which seem less affected by
dense gaseous environments than radio-quiet AGN and which therefore
may be simpler systems on galactic scales. According to the current
paradigm, radio-loud AGN all have relativistically outflowing jets,
and depending on the orientation of the jet, they present markedly
different appearances to the observer (Urry \& Padovani 1995). The
jet luminosity is a critical parameter, influencing the spectral 
energy distributions (Sambruna et al. 1996; Fossati et al. 1997, 1998)
and radio morphological properties (Baum, Zirbel \& O'Dea 1995).
BL~Lac objects have intrinsically lower luminosities than radio-loud quasars
but can be seen to moderately high redshifts thanks to fortuitous
relativistic beaming, hence they form the ideal sample for our study.
BL~Lac nuclei are also relatively weaker than in beamed quasars,
making them even easier targets for host galaxy studies.

Our well-defined survey of more than a hundred BL~Lac objects included the 
full range of BL~Lac types (Padovani \& Giommi 1995), at redshifts
out to $z\sim1.3$. 
The observations and data analysis are described very briefly in
\S~\ref{sec:obsda}, with details given by Scarpa et al. (2000).
Results are given \S~\ref{ssec:results} and discussed further in
\S~\ref{sec:disc}. Conclusions are given in \S~\ref{sec:conc}. For
ease of comparison to the published literature, we used \cosmo
throughout the paper.

\section{Observations and Data Analysis}
\label{sec:obsda}

\subsection{The BL~Lac Sample}
\label{ssec:sample}

The BL~Lac snapshot survey is based on seven complete samples
selected at radio, optical, and X-ray wavelengths (see Table~1). 
We deliberately targeted both radio-selected and X-ray-selected 
BL~Lac objects because the sample content varies strongly with
selection wavelength (Ledden \& O'Dell 1985; Stocke et al. 1985). 
Specifically, there is a selection effect for BL~Lac ``type''
because the two broad components 
in BL~Lac spectral energy distributions (SEDs) 
have peak power outputs ($\nu L_\nu$) at wavelengths that 
increase systematically with luminosity 
(Sambruna, Maraschi \& Urry 1996,;
Fossati et al. 1997,;Ulrich, Maraschi \& Urry 1997).
``Red'' BL~Lacs, also known as low-frequency-peaked BL~Lacs (LBL),
have SEDs peaking at infrared-optical wavelengths and in the
MeV-GeV gamma-ray band, and have luminosities approaching those of quasars.
``Blue'' or high-frequency-peaked BL~Lacs (HBL)
have SEDs peaking at UV-X-ray wavelengths and again at TeV energies,
and are generally less luminous.
Because of these diverse spectral shapes, 
``red'' BL~Lacs dominate existing radio-selected samples
and ``blue'' BL~Lacs dominate most X-ray-selected samples
(Padovani \& Giommi 1995),
although there is clearly a continuous distribution of SED shapes
between these extrema (e.g., Laurent-Muehleisen et al. 1998;
Fossati et al. 1997; Perlman et al. 1998).
Since BL~Lac SEDs are dominated by beamed emission from 
aligned relativistic jets, from radio through gamma-ray wavelengths
(Urry \& Padovani 1995; Ulrich et al. 1997), 
the range of SED shapes sampled in our {\it HST} snapshot survey 
corresponds to the full range of jet physics in BL~Lac objects.

The final list of 132 BL~Lac objects (some are in more than one
sample) was approved for snapshot observations in Cycle~6, 
and in the end 110 were observed. 
These spanned the redshift range $0.027 \leq z \leq 1.34$, 
with a median redshift of $\langle z \rangle =0.29$
and 22 having $z>0.5$; the distribution of redshifts is shown
in Figure~1.
Ten of the 132 were also observed with WFPC2 in Cycle~5, 
for longer exposures in a different filter
(Falomo et al. 1997; Jannuzi et al. 1997; Yanny et al. 1997;
Urry et al. 1999).

\subsection{HST Observations}
\label{ssec:obs}

The list of observed objects is given in Table~2, along with the
redshift and SED type (HBL or LBL). Scarpa et al. (2000) give a more
detailed journal of the observations, including the BL~Lac position,
date of observation, and exposure information. 
Observations were done with the {\it HST} WFPC2 and the F702W filter, 
a sensitive, red, broad-band filter that minimizes contamination from 
extended emission line gas (which is in any case much less important 
in BL~Lacs than in quasars and radio galaxies), dust, and recent star 
formation. In the few cases for which WFPC2 F814W images already existed,
we used the F606W or F555W filters instead, 
to get a broader baseline for estimating colors.
The scheduling was done in snapshot mode, meaning the observations fit
random holes in the {\it HST} schedule for regular GO observations.
Thus, the final list of 110 observed BL~Lac objects constitutes an
unbiased subset of the original target list.

To obtain for each target a final image well exposed
both in the inner, bright nucleus and in the faintest outer regions 
of the host galaxy, we used a series of exposures ranging from a few tens 
of seconds to as long as $\sim 1000$ seconds. From 3 to 5
images were obtained for each target, and were
later combined to remove cosmic-ray events and to increase the signal-to-noise
ratio of the final image. The median exposure times are
480~seconds for $z<0.5$ and 840~seconds for $z>0.5$.

\subsection{Data Reduction and Galaxy Surface Brightness Profile Fitting}
\label{ssec:datared}

Data reduction was carried out as summarized by Urry et al. (1999) and
described in more detail by Scarpa et al. (2000), who also show
the final summed images. Magnitudes reported here are in
the Cousins system, transformed from {\it HST} magnitudes assuming colors
appropriate for a redshifted elliptical galaxy (for details, see
Scarpa et al. 2000). For the host galaxies detected here ($z\lesssim 0.7$), 
the color corrections are $\lesssim 0.6$~mag because the 
Cousin R and WFPC2 F702W filters are similar, 
as are the Johnson V and WFPC2 F606W.

We estimated the expected amount of reddening 
due to interstellar matter in our Galaxy 
from HI column densities, using the
conversion $\log N_H /E(B-V) = 21.83$~cm$^{-2}$ mag$^{-1}$ appropriate
for high latitudes
(Shull \& Van Steenberg 1995), assuming a total-to-selective extinction
$A_R = 2.3 E(B-V)$ (Cardelli, Clayton \& Mathis 1989).
In general, the reddening is
quite small, with median value 0.2~mag at R; 
values for each object are in Table~2.
These corrections were applied to the reported absolute magnitudes,
to give our best estimate of the intrinsic physical quantity,
but not to the apparent magnitudes, which reflect directly measured
quantities. 
In any case, the reddening corrections are usually comparable to or
smaller than the estimated systematic uncertainties (\S~\ref{ssec:errors}).
In three cases $A_R$ is as high as $\sim 2.5$, but for these BL~Lacs
the redshift is not known, so they do not affect our conclusions about
absolute quantities.
Note that we had no constraints on, and therefore did not correct for,
reddening in the host galaxy or BL Lac nucleus itself.

To evaluate the morphologies and apparent magnitudes of the BL~Lac
host galaxies, most of which are quite smooth and round, we fitted
one-dimensional surface brightness profiles. This is computationally
much simpler than two-dimensional analysis and even for our
well-exposed (2-orbit) Cycle~5 images gave equivalent results (Falomo
et al. 1997; Urry et al. 1999); extensive two-dimensional analysis
for the low-redshift ($z<0.3$) BL~Lacs is described by Falomo et al. (2000). 
Azimuthal averaging also improves the
signal-to-noise ratio in the outermost parts of the host galaxy,
allowing us to go to fainter surface brightnesses.

Details of the fitting procedure can be found in Scarpa et
al. (2000), and Urry et al. (1999). Briefly, we fitted the profile
with a galaxy plus point source, convolved with the point spread function (PSF),
adjusting the parameters
simultaneously to determine a best-fit and statistical errors using
the chi-squared statistic. The PSF consisted
of a Tiny Tim model (Krist 1995) in the inner 2~arcsec joined smoothly
to a composite stellar profile for the wings (the pure Tiny Tim PSF
model does not include large angle scattered light, leading
to overestimates of the host galaxy brightness; see Fig.~2 of
Scarpa et al. 2000). 

We tested both exponential disk and de~Vaucouleurs $r^{1/4}$ models
for the galaxy. We used an F-test to evaluate which if either of the two galaxy 
models
was preferred, at 99\% confidence or better. Our
threshold for formal detection of a host galaxy was that
PSF-plus-galaxy fit be better at 99\% confidence than the PSF-only
fit. For unresolved objects we determined 99\% confidence upper
limits (statistical errors) to the host galaxy magnitudes ($\Delta
\chi^2 = 6.6$ for one parameter of interest, $M_{gal}$), fixing the
half-light radius at $r_e =10$~kpc, slightly larger (to be
conservative) than the median ($<r_e>=8.5$~kpc) for the
72 resolved objects.

We did use two-dimensional analysis to test for decentering of the nucleus 
with respect to the host galaxy, for the $\sim30$ well-resolved cases
(Falomo et al. 2000). 
With the exception of peculiar cases like double nuclei (Scarpa et
al. 1999), the point sources are well-centered in the host galaxies,
with a tolerance generally better than 0.03~arcsec. Extensive
two-dimensional analysis carried out on the $\sim30$ nearest host
galaxies ($z<0.3$), where spatial resolution and signal-to-noise ratio
are highest, also shows very small ellipticities (generally $\epsilon
< 0.2$) and few cases of isophotal twists or distortions (analysis and
results described fully by Falomo et al. 2000).

\subsection{Systematic Errors}
\label{ssec:errors}

Comparison of our fitted host galaxy magnitudes with those obtained
by other authors, even on the same data (e.g., Jannuzi et al. 1997)
reveals systematic discrepancies of up to several tenths, even for
bright, easily detected elliptical host galaxies. Sources of
systematic error include uncertainty in the PSF shape, variations in
how the PSF is normalized, and uncertainty in the sky background. We
have done extensive simulations to estimate the size of the systematic
uncertainties in our derived magnitudes, as reported by Scarpa et
al. (2000); here we mention the results.

The PSF shape we have adopted and the fitting procedure recover very
accurately the input parameters in simulated data, with uncertainties in
the total galaxy magnitude of less than $0.15$~mag (Gaussian
half-width of the distribution of measurement minus true value; Scarpa
et al. 2000).
The uncertainty in the sky background is more significant, particularly
when the galaxy is only marginally resolved. From 
the simulations we estimate that the typical systematic
uncertainty in measured host galaxy magnitude is $\pm0.2$~mag when the
point source brightness is within 2~mag of the host galaxy. This
uncertainty of a few tenths is over and above the statistical
uncertainties quoted in Table~2.

We also note that derived galaxy properties reported in the literature
can differ widely because of different calibrations, different
aperture sizes for photometry, and different fitting assumptions, as
well as the usual differences in cosmology. Conversion to absolute
magnitude can introduce further discrepancies because K~corrections in
the literature vary widely (King \& Ellis 1985; Frei \& Gunn 1994; 
Fukugita, Shimasaku \& Ichikawa 1995; Kinney et al. 1996). For elliptical
galaxies, published values differ by 0.1~mag at $z\lesssim0.2$ and as
much as 0.5~mag at $z\sim1$. The range of values for spiral galaxies is
similar or perhaps even larger. The difference between the
K~corrections for E-type and Sb-type spectra is of course much larger
(as much as 2~mag at $z\sim1$). The K~correction values we used are
given in Table~2.

\subsection{Results of Host Galaxy Fits}
\label{ssec:results}

Results of the one-dimensional fitting for de~Vaucouleurs models are
summarized in Table~2, along with the 68\% confidence statistical
uncertainties (in most cases, systematic errors dominate; see
\S~\ref{ssec:errors}). Plots of the radial surface brightness
profiles with best-fit de~Vaucouleurs model and residuals, and of the
chi-squared confidence contours for the two parameters of interest
(host galaxy magnitude and effective radius), are shown by Scarpa et
al. (2000), along with images and discussion of individual sources.

In 72 of the 110 BL~Lac objects, host galaxies are detected.
This is strongly dependent on redshift, since our relatively short
exposures become insensitive to $L^*$ galaxies for $z\gtrsim 0.5$.
Figure \ref{fig:histo} shows detection fraction as a function of
redshift for the observed BL~Lac sample.

Like the host galaxy detection rate, other trends with redshift result
from declining signal-to-noise ratios and/or absolute spatial resolution
with increasing distance of the BL~Lac. 
For example, the absolute magnitude of the nucleus increases with
redshift, as expected in a flux-limited sample (Fig.~\ref{fig:Mnuc_z}).
Note that each BL~Lac type spans nearly the whole redshift range,
although there is redshift segregation because the ``red'' BL~Lacs
are systematically more luminous than the ``blue'' BL~Lacs.

The half-light radius of detected host galaxies increases slightly
with redshift (Fig.~\ref{fig:re_z}), 
corresponding to the larger sizes of more luminous host
galaxies (less luminous host galaxies being harder to detect at high
redshift). The uncertainties in the fitted values of $r_e$ also increase
with redshift. The median value over all redshifts is 
$\langle r_e \rangle = 8.5$~kpc.

Because the flux limits of the input sample were almost entirely unaffected
by optical flux, the host galaxy magnitude is in principle not correlated
with redshift. However, two selection effects affect the magnitude range of 
detected host galaxies. First, faint host galaxies will not be detected
around bright nuclei, and second, very bright host galaxies with weak nuclei 
will have been classified as galaxies rather than AGN. 
Figure~\ref{fig:nuc/host} shows how the ratio of nuclear to host galaxy
luminosity is confined to a relatively narrow range by these two effects. 

\section{Discussion}
\label{sec:disc}

\subsection{Luminosities of the BL~Lac Host Galaxies }
\label{ssec:lum}

Table~2 lists the absolute magnitudes of the host galaxy and nucleus
(i.e., point source) for each observed BL~Lac object, calculated from
the best-fit de~Vaucouleurs model parameters using the K~corrections
listed. The 72 detected hosts are very luminous, round galaxies.
Their median absolute magnitude is $\langle M_R \rangle = -23.7$~mag, 
with a relatively small dispersion of 0.6~mag. These results are
largely in agreement with previous, mostly smaller, surveys of BL~Lac
objects (Abraham, Crawford \& McHardy 1991; 
Stickel, Fried \& K\"{u}hr 1993; Falomo 1996; WSY), 
although for individual objects the differences average
$\sim 1$~mag (see Scarpa et al. 2000). 
Because we probe
higher redshifts on average than these ground-based surveys, it is not
surprising that our detected host galaxies are also somewhat more
luminous on average.

Figure~\ref{fig:Mr_z} shows the distribution of our absolute host
galaxy magnitudes for the 85 BL~Lacs with known redshifts. The
median value (dashed line) is roughly one magnitude brighter than 
$L^*$; $L^*_R =-22.4$~mag at low redshift (Efstathiou, Ellis
\& Peterson 1988, converted from $L^*_B$ assuming $B-R=1.56$). The 
BL~Lac host galaxies are similar in luminosity to brightest cluster
galaxies (Taylor et al. 1996; WSY), $M_R = -23.9$~mag (Thuan \&
Puschell 1989, converted from the H band assuming $R-H=2.5$) or to
Fanaroff-Riley type~I radio galaxies (Ledlow \& Owen 1996; cf. WSY),
which are often found in moderate to rich cluster environments.

At high redshifts, we have many upper limits to the host galaxy
magnitudes. Most are uninteresting because the nuclei are quite bright
but a few are faint, $\gtrsim -23$~mag, indicating at least a few high
redshift BL~Lacs ($z\gtrsim 0.3$) have $L^*$-like host galaxies, like
the lower luminosity hosts at $z\lesssim0.1$.

\subsection{Morphologies of the BL~Lac Host Galaxies }
\label{ssec:morph}

In the vast majority of cases, a de~Vaucouleurs $r^{1/4}$ model fit
the data significantly better than an exponential disk model. In only
14 cases, all with relatively low signal-to-noise ratios, were the
two fits even comparable. (In one case, 0446+449, the disk fit was
unequivocally preferred but there was no point source present,
indicating the identification as a BL~Lac is in error; 
see Scarpa et al. 2000 for a full
discussion of all dubious classifications.)
We are not biased against finding disks, since their surface brightness
would fall off slower than the $r^{1/4}$ profile.
Given our large sample of resolved host galaxies, 
we can say at the 99\% confidence level 
that at most $\sim8$\% can be in disk systems, and
our results are consistent with all BL~Lac objects being found
exclusively in elliptical galaxies. 

Table~2 lists the half-light radii of the host galaxies, in
kiloparsecs, from the best-fit de~Vaucouleurs model. As well as being
luminous, the host galaxies of BL~Lac objects are large, and the
larger galaxies tend to be more luminous.
Figure~\ref{fig:mue_re} shows the relation between half-light radius, $r_e$,
and surface brightness at that radius, $\mu_e$, for the detected BL~Lac 
host galaxies with known redshifts. The data describe a linear
trend such that larger, more luminous galaxies have lower central
surface brightnesses. This has been seen in many samples of
ellipticals, in clusters or out, whether radio-loud or not, and is
basically a projection of the fundamental plane for elliptical
galaxies (Djorgovski \& Davis 1987, Hamabe \& Kormendy 1987). The
best-fit correlation for the BL~Lac host galaxies, 
$\mu_e = (3.9\pm 0.9) \log (r_e/{\rm kpc})  + (17.2\pm0.7)$ mag arcsec$^{-2}$, 
is consistent with those reported for FR~I
radio galaxies (Govoni et al. 2000), bright cluster
ellipticals (BCE; Ledlow \& Owen 1995), and non-cluster ellipticals
(Kormendy 1977). It also agrees well with the slope determined
for other, more powerful AGN (McLure et al. 1999).

The implication is that BL~Lac host galaxies have absolutely normal
elliptical morphologies and are somewhat more luminous than average. 
Because they are very round, there is no obvious 
alignment with the more linear radio structures.
The BL~Lac host galaxies also show normal color profiles 
(Kotilainen, Falomo \& Scarpa 1998; Urry et al. 1999), and
follow quite well a $r^{1/4}$ law, so the nuclear activity
appears to have markedly little effect on the galaxy properties.
The integrated colors, where available, are
consistent with redshifted emission from a passively
evolving elliptical galaxy with an old stellar population
(rest-frame colors $R-I = 0.70$, $V-I= 1.31$) and
imply an initial star formation epoch for the host galaxies
of at least $\sim6$~billion years ago (Bruzual \& Charlot 1993).
Bluer data are required to constrain new star formation, 
to which our WFPC2 F814W images are generally not sensitive
at these low redshifts.

Finally, we note that Figure~\ref{fig:mue_re} includes the 14 
morphologically unclassified host galaxies (for which disk and de Vaucouleurs
fits gave similar chi-squared values). That they fit nicely
into the $\mu_e - r_e$ relation for elliptical hosts supports the
idea that these galaxies are indeed ellipticals.

\subsection{Host Galaxies of ``Red'' and ``Blue'' BL~Lacs}
\label{ssec:redblue}

There are systematic differences between ``red'' and ``blue'' BL~Lac
objects --- in luminosity, redshift, and spectral energy distribution.
The ``blue'' objects have less
luminous nuclei and jets with lower kinetic powers (Celotti, Padovani
\& Ghisellini 1997), and dissipate most of their energy in 
synchrotron radiation from highly relativistic electrons. 
The ``red'' objects, which have systematically higher bolometric
luminosities, are probably redder because the highest energy 
electrons cool quickly by Compton scattering ambient UV and X-ray 
photons to gamma-ray energies, which can dominate the bolometric
output (Ghisellini et al. 1998). 
These two classes of BL~Lac object therefore
reflect two different kinds of jets (probably extrema of a continuous
distribution), which result from different jet formation and/or
evolution.

Despite these strong nuclear trends, we find no differences between 
the host galaxies of ``red'' and ``blue'' BL~Lacs,
either in luminosity or size, confirming the earlier result by WSY
for somewhat fewer objects.
This strongly suggests that nuclear
properties, which can strongly influence jet formation and
propagation, do not have a dramatic effect on large-scale host galaxy 
properties (or vice-versa).

\subsection{Comparison to Radio Galaxies}
\label{ssec:rg}

According to unified schemes (Barthel 1989), BL~Lac objects are
FR~I radio galaxies whose jets are aligned along the line of sight
(Urry \& Padovani 1995). This implies BL~Lac host galaxies should be
statistically indistinguishable from FR~I host galaxies. It has been
suggested that the parent population of BL~Lacs might instead be
FR~IIs, or a subset thereof (Kollgaard 1987; Urry \& Padovani 1995; WSY;
Laing 1994). Our host galaxy study directly tests this alternative
unification hypothesis.

The original division between FR~I and FR~II galaxies was
morphological --- whether hot spots occurred at the inner or outer
edges of the radio source, respectively --- and the excellent
correlation of morphology with radio luminosity was noted at the same
time (Fanaroff \& Riley 1974). For low-frequency radio luminosities
below (above) $P_{178} = 2 \times 10^{25}$~W~Hz~$^{-1}$~sr$^{-1}$,
almost all radio sources were type I (II). This clean separation in
luminosity disappears at higher radio frequencies, where the overlap
can be several decades in radio power.

Owen and Ledlow (1994) showed that FR~I/II division depends on both
radio power and optical luminosity, with a diagonal line dividing FR~Is 
from IIs.
If the observed radio power is a measure of the kinetic power of the
jet, and if optical luminosity correlates with the mass of the host
galaxy, the Fanaroff-Riley division can be explained in (at least) two
ways. In the ``nurture'' scenario, more massive galaxies have a
denser interstellar medium better able to decelerate an outflowing
relativistic jet (Bicknell 1995). For a given jet power, the FR~Is
would be in more luminous galaxies than FR~IIs (i.e., to the right of
the diagonal dividing line); or for a given galaxy mass, 
the FR~Is would have less powerful jets than FR~IIs (i.e., below the diagonal 
line).
In contrast, in one ``nature'' scenario,
FR~Is and FR~IIs are distinguished at birth because the power
delivered to the jet depends on a magnetic switch that essentially
links higher power jets with more massive, spinning black holes (Meier
1999). A correlation between black hole mass and galaxy mass then
leads to the diagonal FRI/II dividing line.

Figure~\ref{fig:pext_opt} shows a new version of the Owen \& Ledlow
(1994) diagram of radio power versus optical magnitude. Because we
plot extended radio power instead of total radio power, and host
galaxy magnitude rather than total magnitude, relativistic beaming of
BL~Lac nuclei has no effect, and thus a direct comparison between the
host galaxies of BL~Lac objects and radio galaxies is possible. 
We took FR~I and II galaxies from the 2~Jy sample (Wall \& Peacock 1985)
because it has similar depth and selection criteria as the 1~Jy BL~Lac
sample (Stickel et al. 1991); morphological classifications are from
Morganti, Killeen \& Tadhunter (1993); values for the extended
radio power are from references listed in Table~2; and we  
restricted all samples to $z<0.5$ to avoid luminosity-redshift biases.
Because the sample selection biases still differ, one cannot compare
the distributions in extended radio power and host galaxy magnitude;
rather, unified populations should simply occupy similar regions in 
Figure~\ref{fig:pext_opt}.

The BL~Lacs overlap extremely well with the FR~I galaxies, with only a
few in the FR~II region.
Similarly, radio-loud quasars (also restricted to $z<0.5$)
lie in the FR~II region of the diagram.
Thus, the present data strongly support the
unification picture with FR~I galaxies constituting the bulk of the parent
population. 

Note that the projection of this plot onto the host galaxy magnitude
axis will give statistically distinguishable distributions for BL~Lacs
and FR~Is; formally, in this one dimension alone, FR~IIs might appear
to be a better match (WSY). This is a misleading approach, however, since it
ignores important information about radio power. As Figure~\ref{fig:pext_opt}
clearly shows, BL~Lac objects are not well matched to FR~II radio galaxies.
Instead, what confuses the one-dimensional approach is that BL~Lacs (so far)
have not been found in host galaxies as luminous as the most luminous
FR~Is, nor have they been found in clusters as rich as those FR~Is.
If the absence of very luminous host galaxies and/or rich cluster
environments is significant (Owen, Ledlow \& Keel 1996), it is possible
that dense intracluster environments or extremely massive host galaxies
completely quench any would-be relativistic jet.

\subsection{Near Environments and Close Companions}
\label{ssec:near}

Further refinement of the unification picture is possible from
consideration of the larger environments of the BL~Lacs (Fried et
al. 1993; Falomo, Pesce, \& Treves 1993, 1995; Pesce, Falomo, \&
Treves 1994, 1995; Smith, O'Dea \& Baum 1995; Wurtz et al. 1997).
Since FR~I radio galaxies commonly occur in clusters, so should BL~Lac
objects. The present data add to previous work by allowing detection
of fainter companions closer to the BL~Lac nucleus. The small field of
view, however, limits the statistics with which the possible excess of
companion galaxies can be assessed.

Preliminary results for the environments of BL~Lacs indicate a large 
number of objects with companions, some as close as 5~kpc (projected).
Defining ``companion'' galaxies as 
those within 70~kpc of the BL~Lac nucleus and brighter than 
1~magnitude below $m^*$ (for those objects without measured redshift, 
$z=0.2$ was used), we find companion galaxies in 42\% of the 
BL~Lac sample. Without the magnitude limit, so that fainter galaxies
are included, 47\% of the sample has companions.
For comparison, 42\% of a sample of low redshift FR~I 
galaxies have companion galaxies within the same radius (Pesce et al.,
in preparation.).

On the larger scale environment, BL~Lacs have been seen to lie in 
regions of enhanced galaxy density, on average. Typically, 
the clusters around BL~Lacs are poor, of Abell richness class 0-1, 
although a few richer clusters are detected. Our preliminary results 
for the snapshot survey are similar, with 40\% of the sample showing 
regions of enhanced galaxy density. Nonetheless, a 
significant number of objects appear to be completely 
isolated (i.e., no close companions and no surrounding galaxies 
above the average background).
A comprehensive analysis of the environments of BL~Lac objects,
determined from the {\it HST} images, will be given in separate papers
(Pesce et al. and Falomo et al., in preparation).

If mergers are a significant part of the galaxy formation process
(especially ellipticals) then hosts should be more disturbed at high
redshift. The BL~Lac objects in our sample generally appear
undisturbed, with a few exceptions (details of individual sources are
given by Scarpa et al. 1999, 2000).
This is in marked contrast to the case for more powerful radio sources.
Since among BL~Lacs, jet power has no discernible effect on galaxy morphology,
this suggests that instead age may be important.
Radio sources having undergone recent mergers would be more likely to show
dust lanes, tidal tails, and the like, even if most of the stars were formed
at high redshift. In contrast, the relaxed morphologies
of BL~Lac host galaxies suggest they are old, more evolved sources
that have not recently merged.
The observation that radio-quiet AGN have less disturbed morphologies 
than radio-loud AGN (Hutchings et al. 1989) could then
be a function more of intrinsic AGN luminosity and/or evolutionary
state than radio-loudness, since our radio-loud sample is markedly undisturbed.
Comparison to observations of BL~Lac objects at higher redshifts ($z>0.5$)
will be very illuminating on this point.

\subsection{Comparison of Nucleus and Host Galaxy }
\label{ssec:nuc_hg}

A correlation or trend between galaxy magnitude and nuclear brightness has been
reported in several host galaxy studies (McLeod \& Rieke 1994, 1995; 
Hooper et al. 1997; McLeod et al 1999). This can be interpreted as an 
extension of the
correlation between black hole mass and bulge mass in nearby
ellipticals (Kormendy \& Richstone 1995; Magorrian et al. 1998; 
van der Marel 1999), providing the
Eddington ratio does not vary widely among the AGN considered
(cf. McLure et al. 1999).

In our sample of BL~Lac objects, there is a slight correlation
between measured nuclear and host galaxy luminosities, 
but it becomes insignificant when upper limits are included. 
This can be seen in Figure~\ref{fig:hg_nuc}, 
which shows that the host galaxy magnitudes
cluster near the median value, $\langle M_R \rangle = -23.7$~mag,
independent of the luminosity of the nucleus. 
Furthermore, the best-fit slope of a linear relation 
is much shallower than implied by the correlation between black hole mass and 
bulge
mass for fixed Eddington ratio. If the bulge-black hole correlation
translates to $L_{gal}$-$L_{nuc}$ in our data, then the Eddington ratio
must range over at least two orders of magnitude among otherwise
similar jet sources.

The observed point source magnitude, which is dominated by synchrotron
emission from an unresolved jet, is likely affected by relativistic
beaming. This could cause the points plotted in Figure~\ref{fig:hg_nuc}
to extend across an artificially large range in point source magnitude, 
possibly washing out an underlying correlation. 
Estimates for the Doppler factor (actually lower limits)
are available for only a fraction of our target sources 
(Burbidge \& Hewitt 1987; Xie et al. 1991; Dondi \& Ghisellini 1995), 
so wholesale correction of
the observed nuclear magnitudes is not presently feasible.
Instead we considered whether extended radio power --- which 
correlates with jet power and is unaffected by beaming ---
was correlated with host galaxy magnitude. Including upper limits,
there is no significant correlation; ignoring upper limits, there is
a marginal correlation, with a slope much shallower than that implied
by the bulge-black hole relation.

It remains to be explained why a nucleus-galaxy correlation is seen in some 
other
samples and not in the present sample of BL~Lac objects. One
possibility is that there is a luminosity threshold for the effect and
that it does not appear in low-luminosity AGN, as suggested by 
McLeod \& Rieke (1995). Our sample includes
some luminous AGN at $M_R < -25$~mag, the region where
McLeod \& Rieke (1995) found a correlation in their quasar sample, 
but few have detected host galaxies and the nuclear magnitudes are
affected by beaming.
Combining our BL~Lac sample with quasar samples matched in redshift,
it should be possible to assess this issue directly.

Two selection effects could in
principle induce a spurious correlation, especially in data with the
low spatial resolution typical of ground-based observations: (1) the
difficulty of finding faint host galaxies around bright nuclei and (2)
the absence of AGN with bright host galaxies and weak nuclei (these
are identified as galaxies rather than AGN). For any given
investigation, simulations can indicate whether these effects are
significant. (We note that because McLeod \& Rieke 1995 detected host
galaxies for 100\% of the AGN in their sample, the correlation they 
report should not be influenced by the first effect.) 

The correlation of luminosity with redshift in flux-limited samples
could confuse the effects of evolution or steep luminosity functions
with physical effects like true nuclear/host galaxy relations.
To measure the latter effect definitively therefore require spanning 
a large range of luminosity at a fixed redshift.
At present, conclusions drawn from flux-limited samples of limited 
luminosity range at any one redshift, whether high luminosity (quasars) 
or low-luminosity (BL~Lacs), must be considered tentative. 

Comparing the positions of nucleus and host galaxy, we are able, 
with the high spatial resolution of {\it HST}, to place
tight limits on any de-centering of the BL~Lac nucleus. If any of our
BL~Lac objects were actually background quasars microlensed by stars
in a foreground galaxy (which we are calling the host galaxy), there
could well be an offset between the position of the nucleus (the
amplified background quasar) and the lensing galaxy (Ostriker \&
Vietri 1985). Instead, we find that the nuclei are generally
well-centered in the host galaxy, with deviations typically less than
0.03~arcsec (Falomo et al. 1998, 2000). Thus there is no evidence
for microlensing occurring in a large fraction of our sample.

\section{Conclusions}
\label{sec:conc}

We have shown that with {\it HST} it is easy to detect and characterize
the host galaxies of low-luminosity AGN like BL~Lac objects,
up to moderately high redshifts, $z \sim 0.6$. 
We detected host galaxies in almost all cases with $z<0.5$
(58 of 63), and in 6 of 22 with $z>0.5$.
The highest redshift BL~Lac object with a detected host galaxy
is 1823+568 at $z=0.664$ (Falomo et al. 1997).

The detected host galaxies are smooth, round, very luminous ellipticals,
well fitted with de~Vaucouleurs surface brightness profiles. 
In most cases --- generally, where the signal-to-noise ratio 
is high --- the $r^{1/4}$ law fits significantly better than an
exponential disk; in the remaining cases, neither fit is preferred.
Thus our data are consistent with all BL~Lac host galaxies being
ellipticals. 

The median K-corrected absolute magnitude of the detected host
galaxies is $\langle M_R \rangle = -23.7$~mag, with a dispersion of
0.6~mag. This is more than 1~mag brighter than $L^*_R$ galaxy, and
comparable to brightest cluster galaxies, or to
Fanaroff-Riley type~I radio galaxies (Ledlow \& Owen 1996) which are
often found in moderate to rich cluster environments. This strongly
supports the unification of BL~Lac objects with low-luminosity radio
galaxies, and rules out the possibility, at least at these low
redshifts, that the parent population of a substantial fraction of 
BL~Lacs is FR~II radio galaxies. Note that there is a decade or so of overlap
between the FR~I and II populations, and between the quasar and BL~Lac
populations, not quite the clean division that was the original paradigm.

The BL~Lac host galaxies follow the same trend in
the $\mu_e$-$r_e$ projection of the fundamental plane as
other luminous elliptical galaxies. By any measure, BL~Lac
host galaxies look like completely normal ellipticals that are
somewhat brighter than average --- as far as the galaxy goes,
there is no evidence of the nuclear activity.

There are no systematic differences in the host galaxies of ``red'' and
``blue'' BL~Lac objects, once the obvious selection effects (on the BL~Lac
nuclei) are taken into account. Thus active nuclei with relativistic jets of 
very different kinetic powers can live in very similar galaxies.
Their formation cannot be strongly affected by galaxy mass or morphology,
nor can their effect on the host galaxy be dramatic.

We confirm previous studies that BL~Lacs tend to lie in regions of
enhanced galaxy density, either groups or poor clusters, although the
small WFPC2 field-of-view limits the statistical significance of this
result. In some cases, however, the BL~Lac object appears truly
isolated, with no nearby companions or surrounding cluster galaxies, 
to limits several magnitudes below the BL~Lac brightness.

We can rule out that a substantial fraction of BL~Lac objects at (apparently)
low redshift are actually high-redshift quasars microlensed by 
intervening galaxies. Were the detected galaxies not hosts but 
lensing galaxies,
in at least some cases the nuclei should be displaced from the center of
the galaxy. With our very large sample, we can say with high confidence
that this is not the case.

Contrary to previous studies, we do not find any correlation between nuclear
and host galaxy luminosities, such as might have been expected from the
trend of black hole mass with bulge mass seen in nearby ellipticals.
Although the observed nuclear properties of the BL~Lac objects are clearly
affected by beaming, correction for this effect makes no difference to
the lack of correlation. 
Simulations show that selection effects, wherein bright nuclei can obscure
all but the brightest host galaxies, could contribute to spurious correlation.
The lack of an observed correlation for low-luminosity radio-loud AGN 
implies a large scatter in Eddington ratio.

The unification of radio-loud AGN is strongly supported by our results.
This means that the properties of BL~Lac host galaxies and near environments 
are basically universal to all low-luminosity radio-loud AGN. Just as 
FR~I and FR~II radio galaxies span the full range of central engine power,
so do BL~Lacs represent the low-luminosity version of radio-loud quasars.
To understand fully trends in luminosity and/or redshift, samples of
BL~Lac objects and quasars should be combined, as we intend to do in 
future work.

\acknowledgements
Support for this work was provided by NASA through grant number GO6363.01-95A
from the Space Telescope Science Institute, which is operated
by AURA, Inc., under NASA contract NAS~5-26555. MG acknowledges support from
the Hubble Fellowship Program through grant number HF 1071.01-94A, awarded by
the Space Telescope Science Institute. We thank Marcella Carollo, 
Bob Hanisch, John Hutchings, Marek Kukula, and Susan Ridgway
for useful and stimulating discussions, 
and Jochen Heidt, Kim McLeod, Ross McLure, and Chris O'Dea 
for helpful comments on the manuscript. 
RF thanks the {\it HST} visitor program for hospitality
during several visits to STScI. We thank G. Fasano for the use of the AIAP
package. This research made use of the NASA/IPAC Extragalactic Database (NED),
operated by the Jet Propulsion Laboratory, Caltech, under contract with NASA,
and of NASA's Astrophysics Data System Abstract Service (ADS).

\newpage
\leftline{Figure Captions}
\bigskip

\figcaption[]{Histogram of redshifts for the observed BL~Lac
objects. Those with resolved host galaxies are indicated by 
cross-hatching. Relatively few host galaxies are detected for $z>0.5$, 
only 6 of 22, due to the relatively short snapshot exposures.
For $z<0.5$, in contrast, 92\% of the host galaxies are detected.
Among those objects with unknown redshifts (shown in the bin at $z<0$),
only 1/3 have resolved host galaxies, consistent with most being 
at relatively high redshift.
\label{fig:histo}
}

\figcaption[]{Absolute nuclear R magnitude of the observed BL~Lac objects
increases with redshift because the samples are flux-limited 
({\it filled triangles:} ``red'' BL~Lacs, or LBL; 
{\it filled circles:} ``blue'' BL~Lacs, or HBL).
The same is true for radio galaxies ({\it open squares};
Govoni et al. 2000, Chiaberge et al. 1999), which have 
lower nuclear luminosities than BL~Lacs because their jets 
are more nearly in the plane of the sky.
According to unified schemes, BL~Lacs offer an opportunity to study
low-luminosity radio galaxies at higher redshift (Urry \& Padovani 1995). 
\label{fig:Mnuc_z}
}

\figcaption[]{Host galaxy half-light radius versus redshift. The
measured values increase slightly with redshift, corresponding to the 
systematically larger sizes of more luminous host galaxies,
less luminous host galaxies being harder to detect at high redshift. 
There is no difference in the sizes of ``red'' BL~Lacs 
(LBL; {\it filled triangles}) and ``blue'' BL~Lacs (HBL; {\it filled circles}).
\label{fig:re_z}
}

\figcaption[]{The distribution of the observed nuclear-to-host-galaxy luminosity 
ratio is relatively narrow because faint host galaxies are too difficult 
to detect around luminous nuclei, and luminous host galaxies with faint 
nuclei would be classified as galaxies. 
{\it Lower panel:} The histogram of the distribution has a width of
about 2 decades for resolved objects ({\it solid line}),
somewhat broader including unresolved objects ({\it dotted line}; assuming
median host galaxy brightness). 
{\it Upper panel:} The same ratio as a function of redshift.
{\it Filled triangles:} ``red'' BL~Lacs (LBL);
{\it filled circles:} ``blue'' BL~Lacs (HBL);
{\it arrows:} lower limits for unresolved host galaxies, 
with the tip of the arrow corresponding to
a host galaxy one magnitude fainter than the median value $M_R = -23.7$~mag.
The ratio appears to increase with redshift because the nuclear brightness
is increasing (a selection effect) 
while the galaxy magnitudes are essentially constant.
\label{fig:nuc/host}
}

\figcaption[]{K-corrected absolute R~magnitudes of the host galaxies of
the 85 BL~Lacs with known redshifts. 
The median value for detected host
galaxies is $\langle M_R \rangle = -23.7$~mag ({\it dashed line}), 
with a relatively small dispersion
about this value, $\pm0.6$~mag. This is nearly one magnitude brighter than
$L^*$, comparable to brightest cluster galaxies and to
Fanaroff-Riley type~I radio galaxies.
{\it Solid line:} Rest-frame absolute R-band magnitude for a passively evolving
elliptical galaxy with $M_R = -23.7$~mag at $z=0$ (according to the model of
Bressan et al. 1994).
{\it Filled triangles:} ``red'' BL~Lacs (LBL);
{\it filled circles:} ``blue'' BL~Lacs (HBL);
upper limits are shown for unresolved objects.
\label{fig:Mr_z}
}

\figcaption[]{The relation between surface brightness 
at the half-light radius ($\mu_e$) 
and half-light radius ($r_e$) for BL~Lac host galaxies 
(detections with known redshifts only).
{\it Filled triangles:} ``red'' BL~Lacs (LBL);
{\it filled circles:} ``blue'' BL~Lacs (HBL).
The data follow the usual projection of the fundamental plane 
(Djorgovski \& Davis 1987, Hamabe \& Kormendy 1987),
with larger, more luminous galaxies having smaller $\mu_e$.
Similar trends have been found for 
brightest cluster ellipticals ({\it solid line}; BCE, Ledlow \& Owen 1995),
non-cluster ellipticals ({\it dot-dash line}; Hamabe \& Kormendy 1987), 
and radio galaxies ({\it dashed lines}; Govoni et al. 2000).
\label{fig:mue_re}
}

\figcaption[]{Extended radio power versus host galaxy R-band magnitude for
the observed BL~Lacs, along with samples of quasars and radio galaxies
(after Owen \& Ledlow 1994). Fanaroff-Riley type~I ({\it `1' symbols}) 
and type~II ({\it `2' symbols}) radio galaxies are separated 
approximately along a diagonal line in this figure. 
The BL~Lacs 
({\it filled triangles:} ``red'' BL~Lacs [LBL];
{\it filled circles:} ``blue'' BL~Lacs [HBL])
overlap extremely well with the 
FR~I galaxies, with only a few near the FR~II region, while
quasars ({\it stars}) lie in the FR~II region of the diagram.
This figure is unaffected by beaming since we plot extended radio power
instead of total radio power, and host galaxy magnitude rather than
total magnitude.
The BL Lac data are from Table~2 (see references there).
The FR~Is and IIs shown are from the 2~Jy sample (Wall \& Peacock 1985), with
morphological classifications from Morganti et al. (1993).
The quasar data 
are from Taylor et al. (1996), Bahcall et al. (1997), Boyce et al. (1998), 
and Hutchings, Janson \& Neff (1989).
The lines dividing FR~I and FR~II sources
are from the models of Bicknell (1995; see paper for details) and 
represent the extremes of the parameter space he explored:
{\it solid line:} ratio of electron Lorentz factors
$\gamma_{\rm max}/\gamma_{\rm min} = 10^4$, 
synchrotron high-frequency cutoff $\nu_{\rm c} = 10^{10}$~Hz,
and no energy in cold protons ($f=1$);
{\it dotted line:}
$\gamma_{\rm max}/\gamma_{\rm min} = 10^4$, 
synchrotron high-frequency cutoff $\nu_{\rm c} = 10^{11}$~Hz,
and equal energy in electrons and protons ($f=0.5$).
\label{fig:pext_opt} 
}

\figcaption[]{Absolute magnitudes of host galaxy versus nuclear point
source for the 85 BL~Lac objects with known redshifts.
Taking upper limits into account, there is no
significant correlation between host galaxy and nuclear intensity.
The host galaxy magnitudes are narrowly distributed around the median
value, $\langle M_R \rangle = -23.7$~mag, regardless of the luminosity of 
the nucleus.
{\it Filled triangles:} ``red'' BL~Lacs (LBL);
{\it filled circles:} ``blue'' BL~Lacs (HBL).
The relation between black hole mass and bulge
mass found for nearby ellipticals (Magorrian et al. 1998) is transformed
to one between host galaxy magnitude and nuclear magnitude
assuming a mass-to-light ratio $\tau_R = 4$ (as in McLeod et al.  1999),
and Eddington ratios $L/L_{\rm Edd} = 1.0$ ({\it solid line}),
0.1 ({\it dashed line}), and 0.01 ({\it dotted line}).
Radio galaxies ({\it open squares}) with much lower nuclear magnitudes
have similar host galaxy magnitudes 
(Govoni et al. 2000, Chiaberge et al. 1999).
The formal separation between radio galaxies and BL~Lacs 
({\it dot-dash line}) comes from the (arbitrary) classification criterion 
for BL~Lacs that the contrast of the 
4000\AA\ break must be smaller than 25\% (Dressler \& Shectman 1987;
Stocke et al. 1991; Owen et al. 1996). In the R band, this limit 
means that AGN with $m_{host}<m_{nucleus}+1.3$ are
classified as radio galaxies. 
\label{fig:hg_nuc}
}

\newpage
\begin{figure}
  \centerline{
    \psfig{file=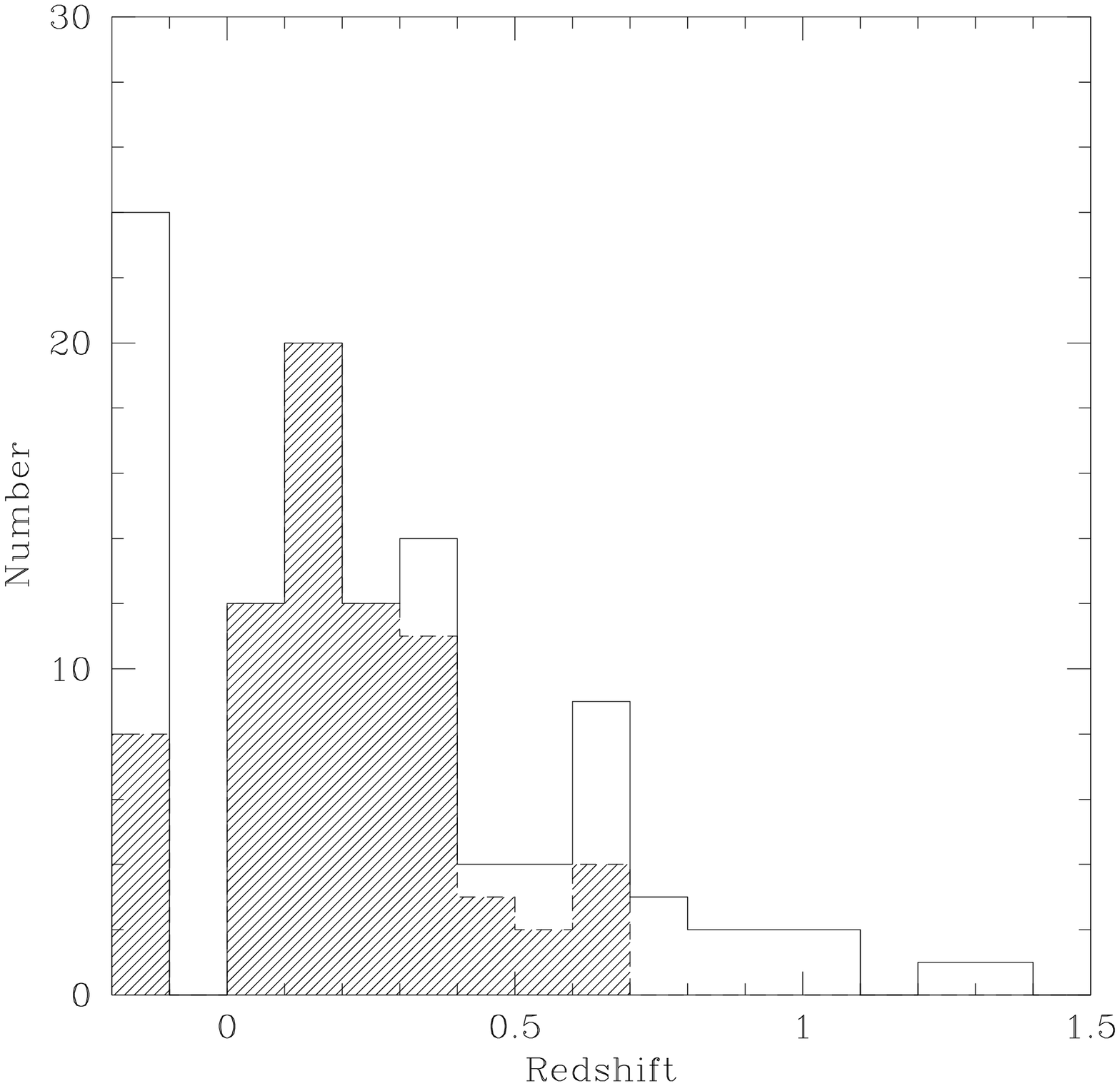,width=0.9 \linewidth}
Fig. 1
  }
\end{figure}
\newpage
\begin{figure}
  \centerline{
    \psfig{file=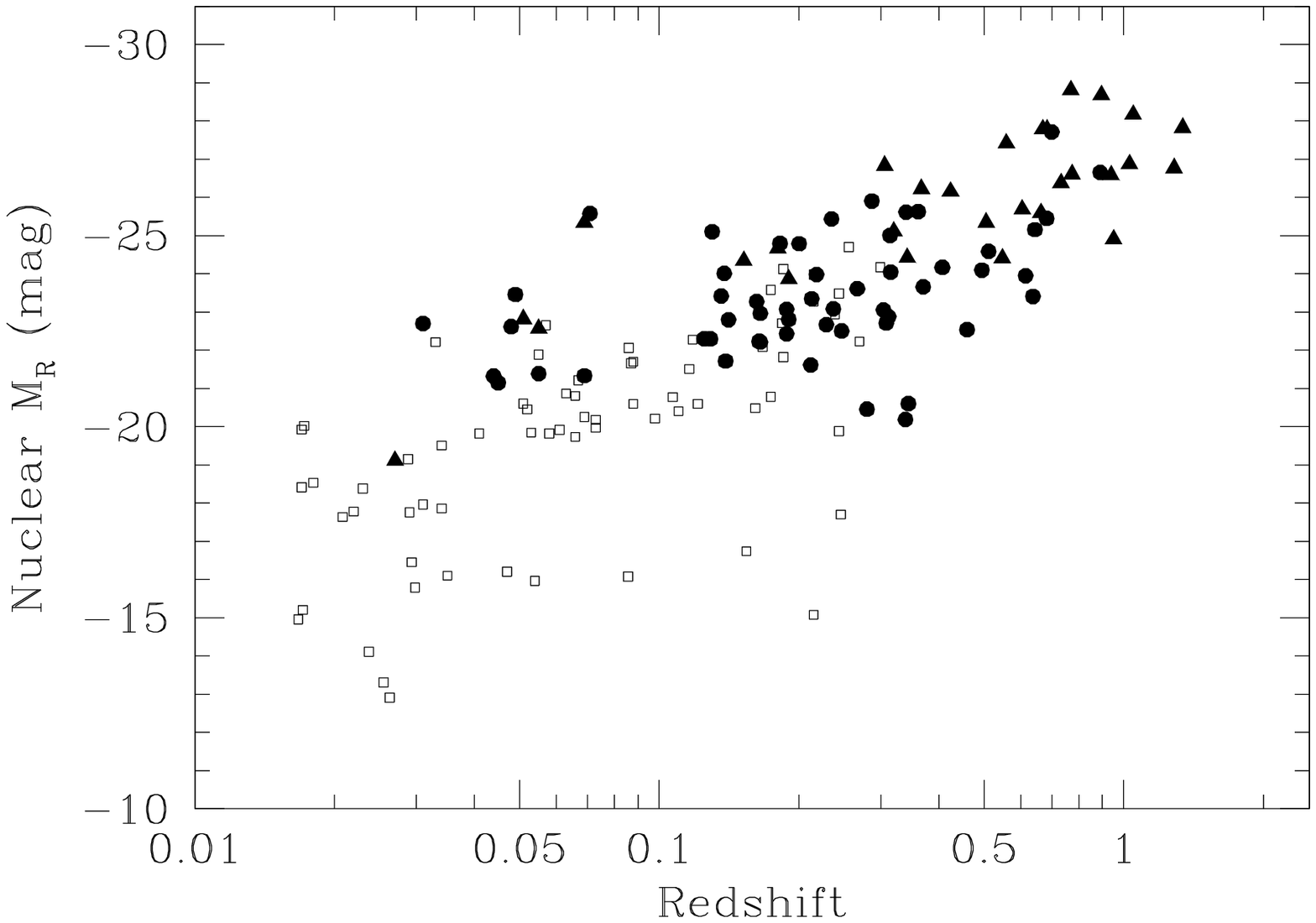,width=0.9 \linewidth}
Fig. 2
  }
\end{figure}
\newpage
\begin{figure}
  \centerline{
    \psfig{file=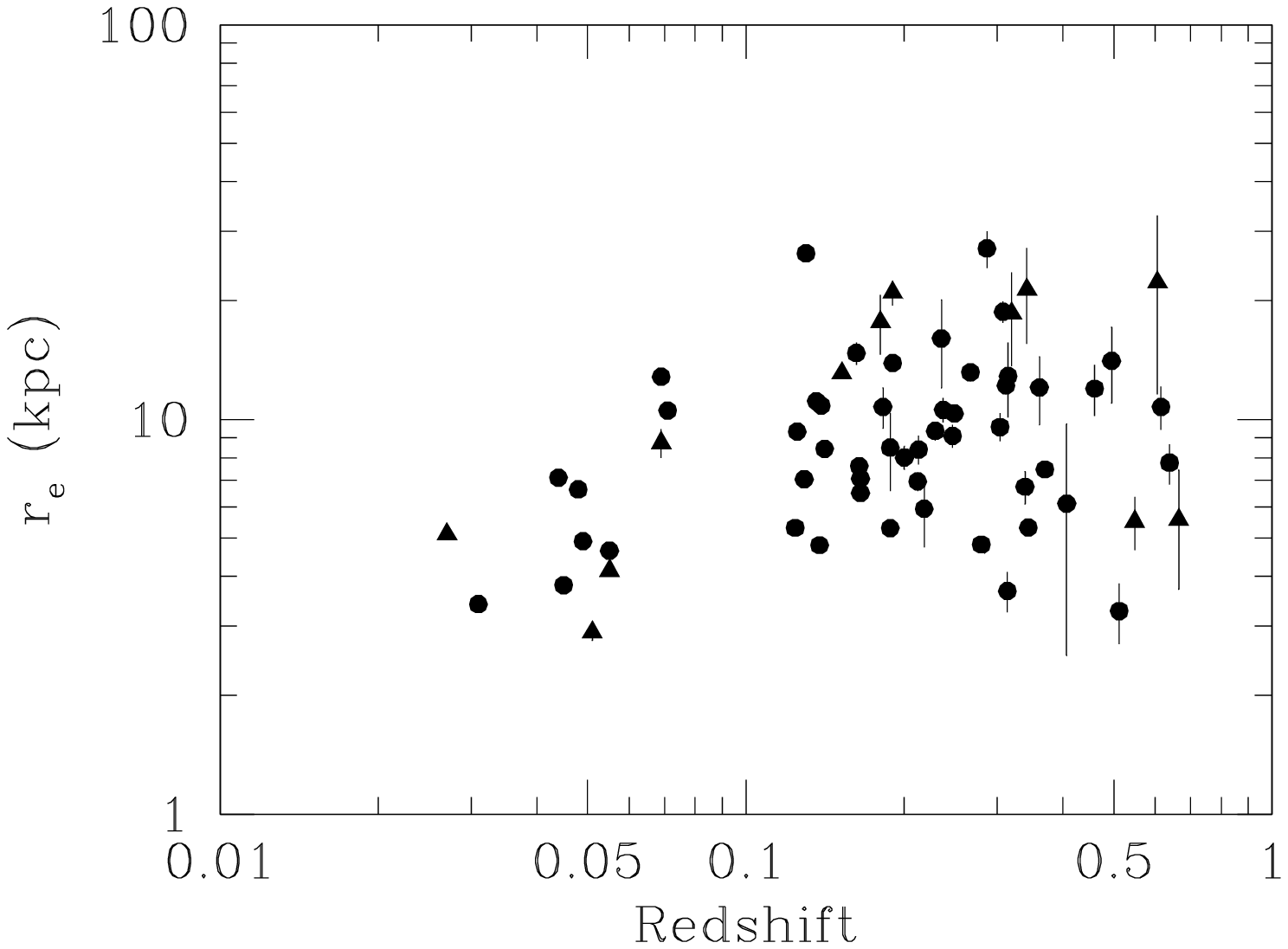,width=0.9 \linewidth}
Fig. 3
  }
\end{figure}
\newpage
\begin{figure}
  \centerline{
    \psfig{file=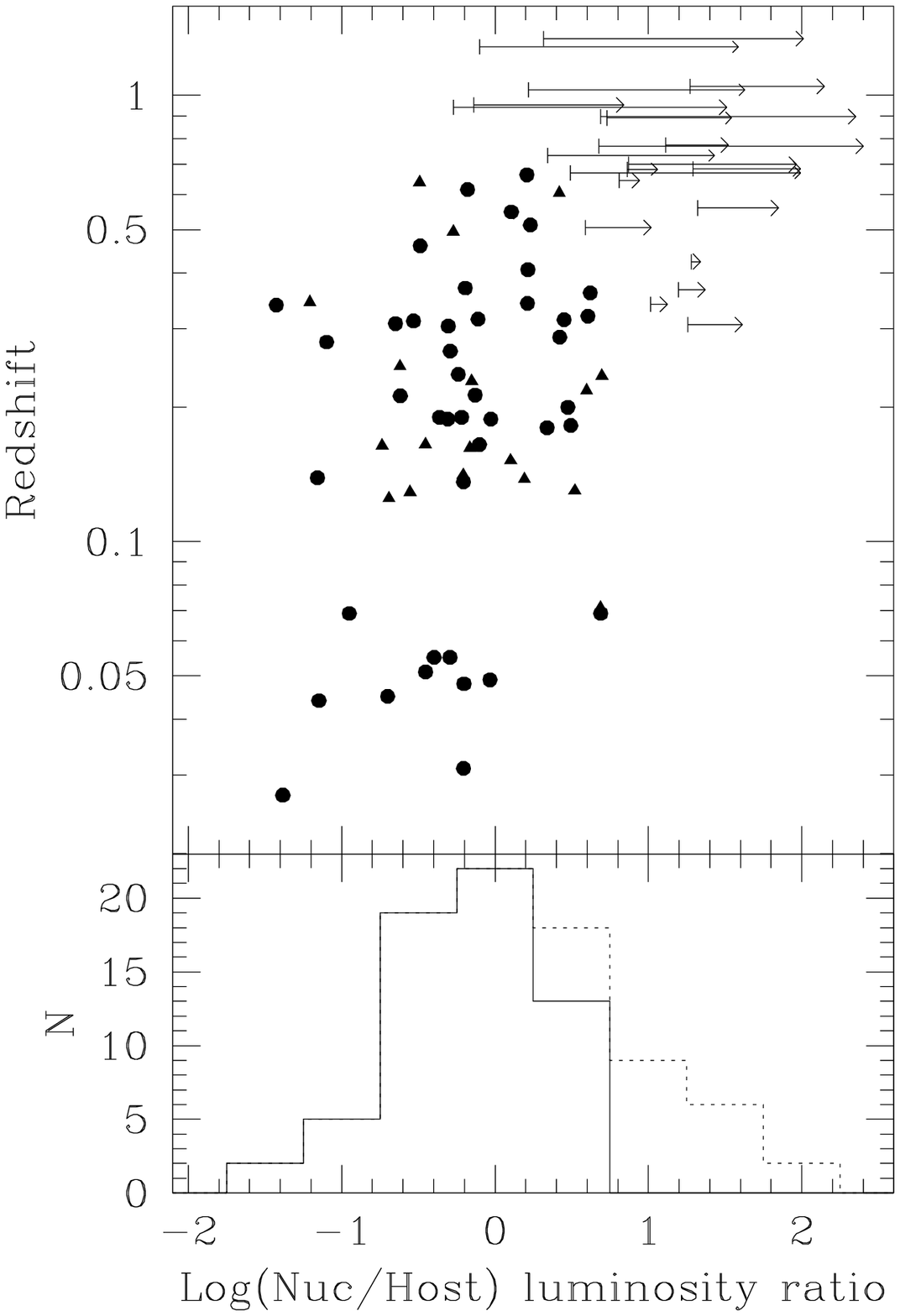,width=0.9 \linewidth}
Fig. 4
  }
\end{figure}
\newpage
\begin{figure}
  \centerline{
    \psfig{file=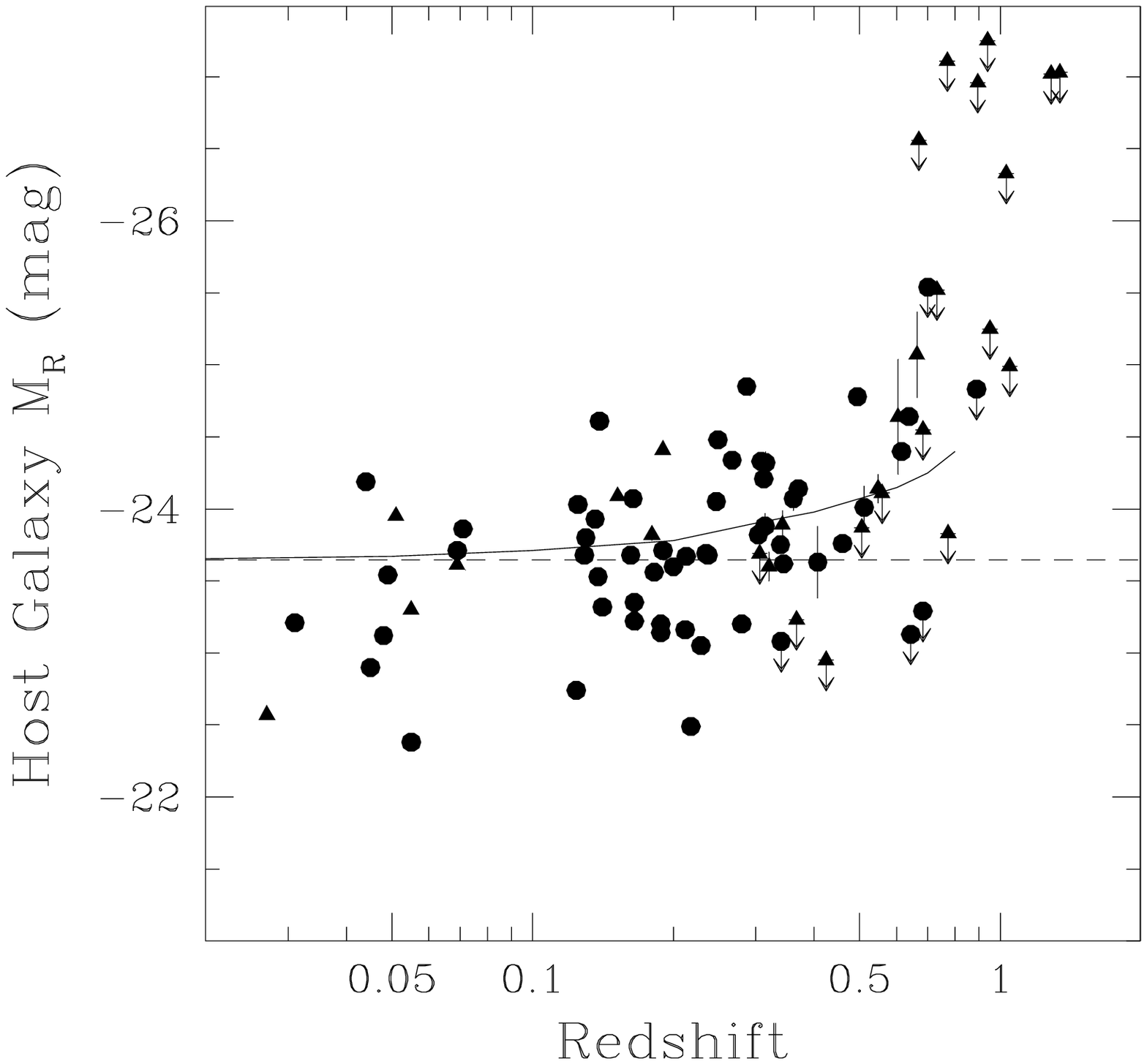,width=0.9 \linewidth}
Fig. 5
  }
\end{figure}
\newpage
\begin{figure}
  \centerline{
    \psfig{file=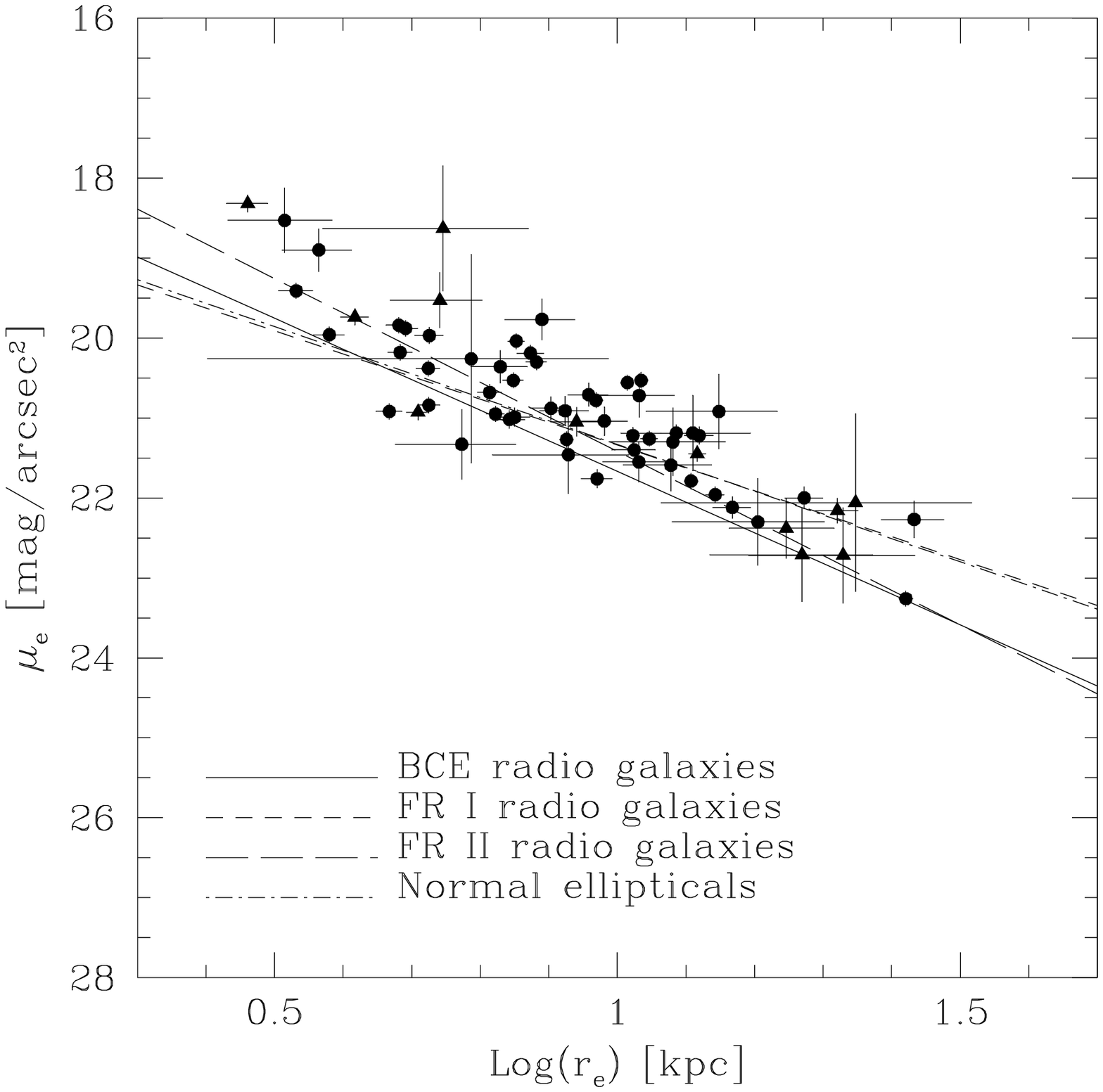,width=0.9 \linewidth}
Fig. 6
  }
\end{figure}
\newpage
\begin{figure}
  \centerline{
    \psfig{file=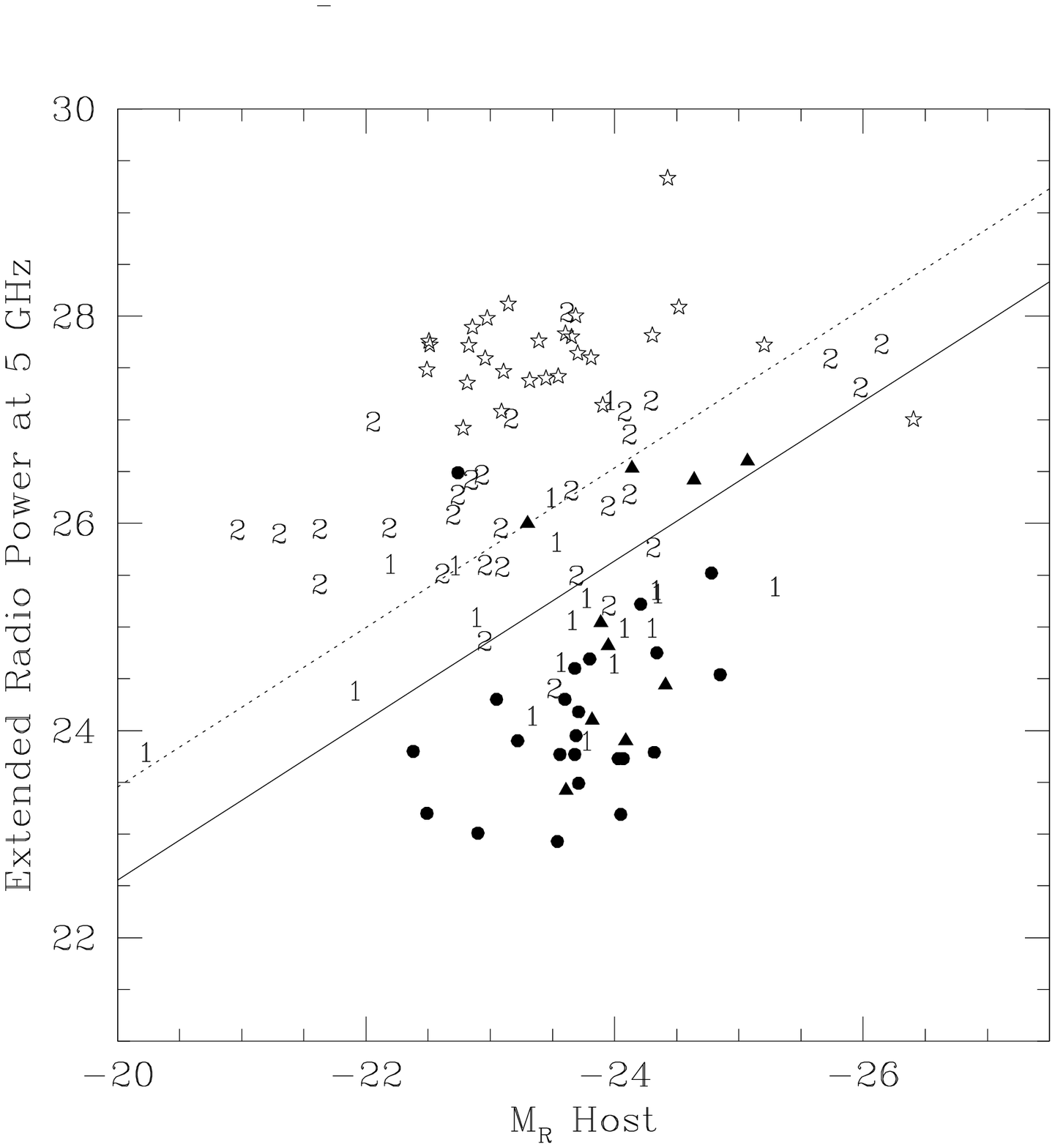,width=0.9 \linewidth}
Fig. 7
  }
\end{figure}
\newpage
\begin{figure}
  \centerline{
    \psfig{file=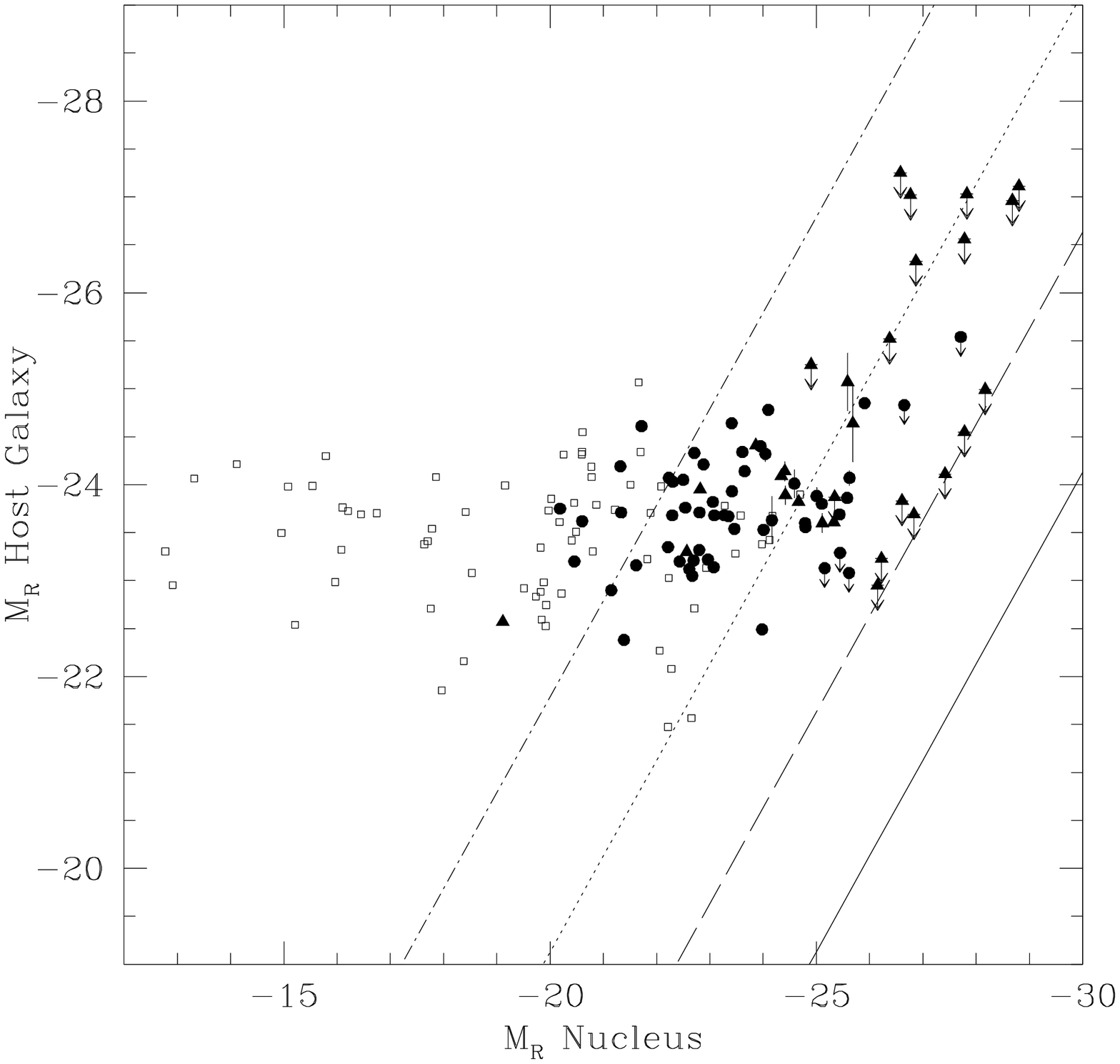,width=0.9 \linewidth}
Fig. 8
  }
\end{figure}

\newpage
\begin{deluxetable}{lccl}
\def\en{\enspace}

\tablenum{1}
\tablewidth{4in}
\tablecaption{BL Lac Samples}
\tablehead{
Sample & N tot & N obs & Reference
}
\startdata
1Jy      & 34  & 30 & Stickel {\it et al.} 1991 \nl
S4       & 14  &  3 & Stickel \& K\"uhr 1994 \nl
PG       & 7   &  6 & Green {\it et al.} 1986   \nl
HEAO-A2  & 6   &  2 & Piccinotti {\it et al.} 1982 \nl
HEAO-A3  & 27  & 22 & Remillard {\it et al.} 1994 \nl
EMSS     & 36  & 23 & Morris {\it et al.} 1991 \nl
SLEW     & 28  & 23 & Schachter {\it et al.} 1993 \nl
         &     &    & Perlman {\it et al.} 1996 \nl
\enddata
\end{deluxetable}

\begin{figure}
   \vspace{-3cm}
  \centerline{
    \psfig{file=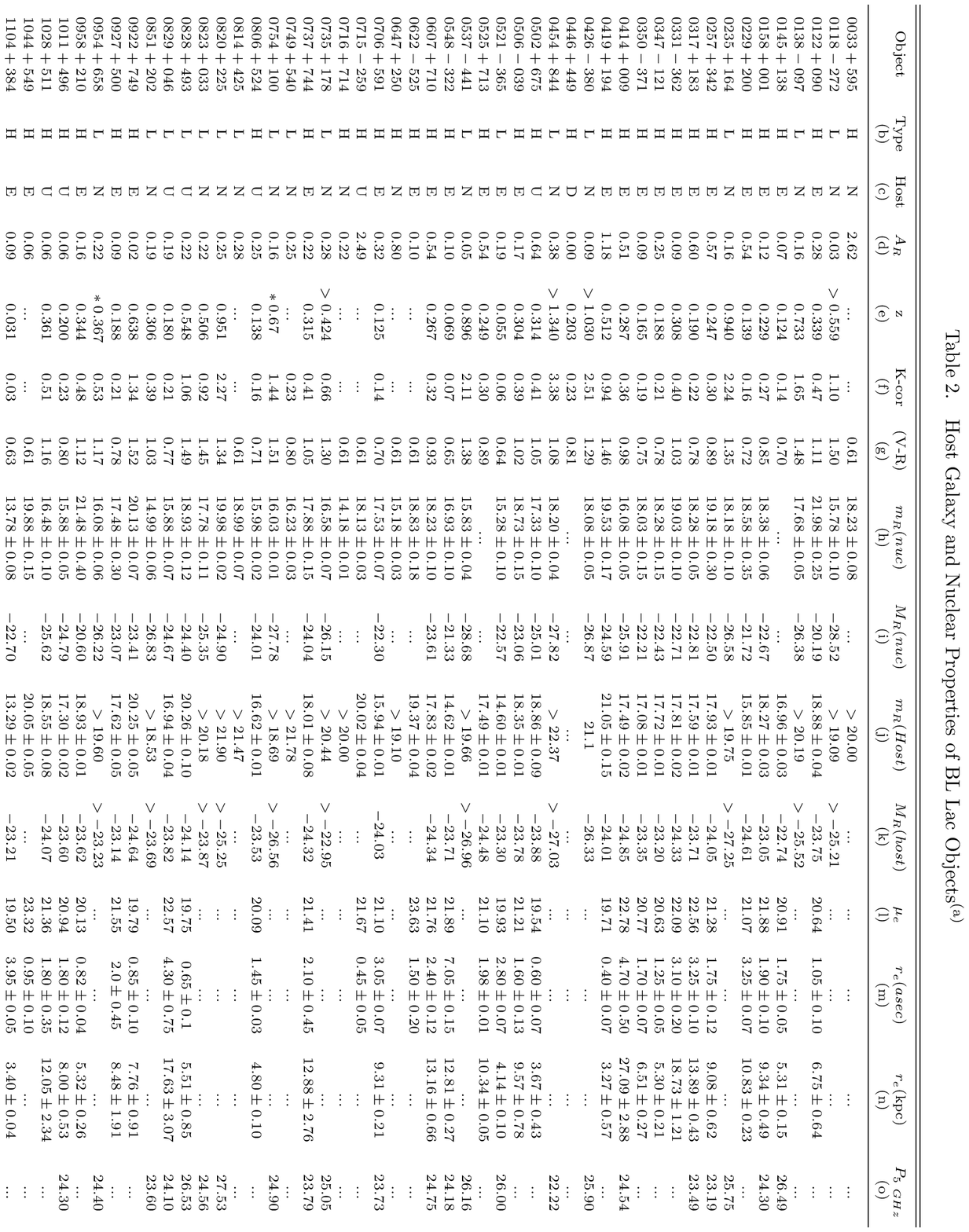}
  }
\end{figure}

\begin{figure}
   \vspace{-3cm}
  \centerline{
    \psfig{file=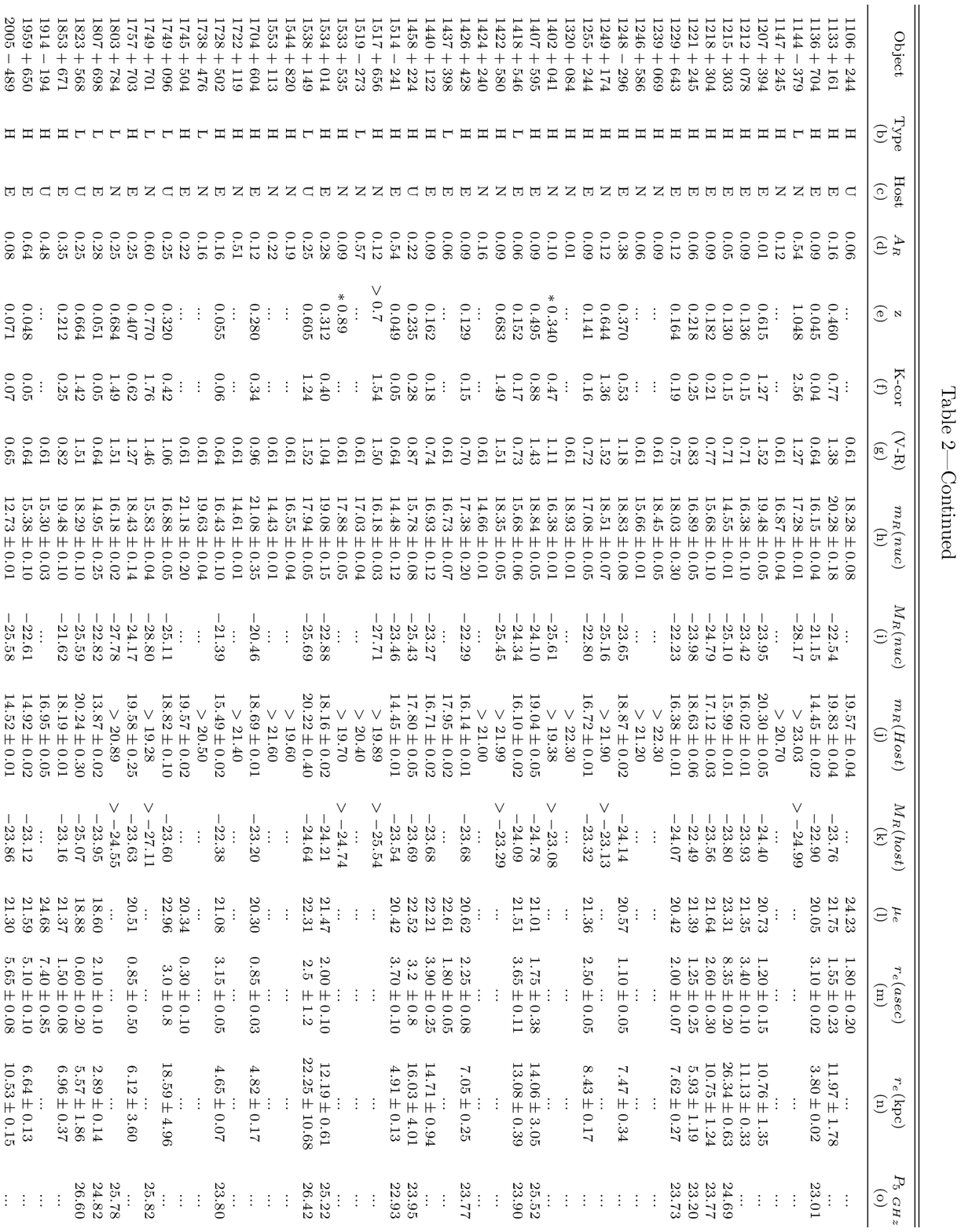}
  }
\end{figure}

\begin{figure}
   \vspace{-3cm}
  \centerline{
    \psfig{file=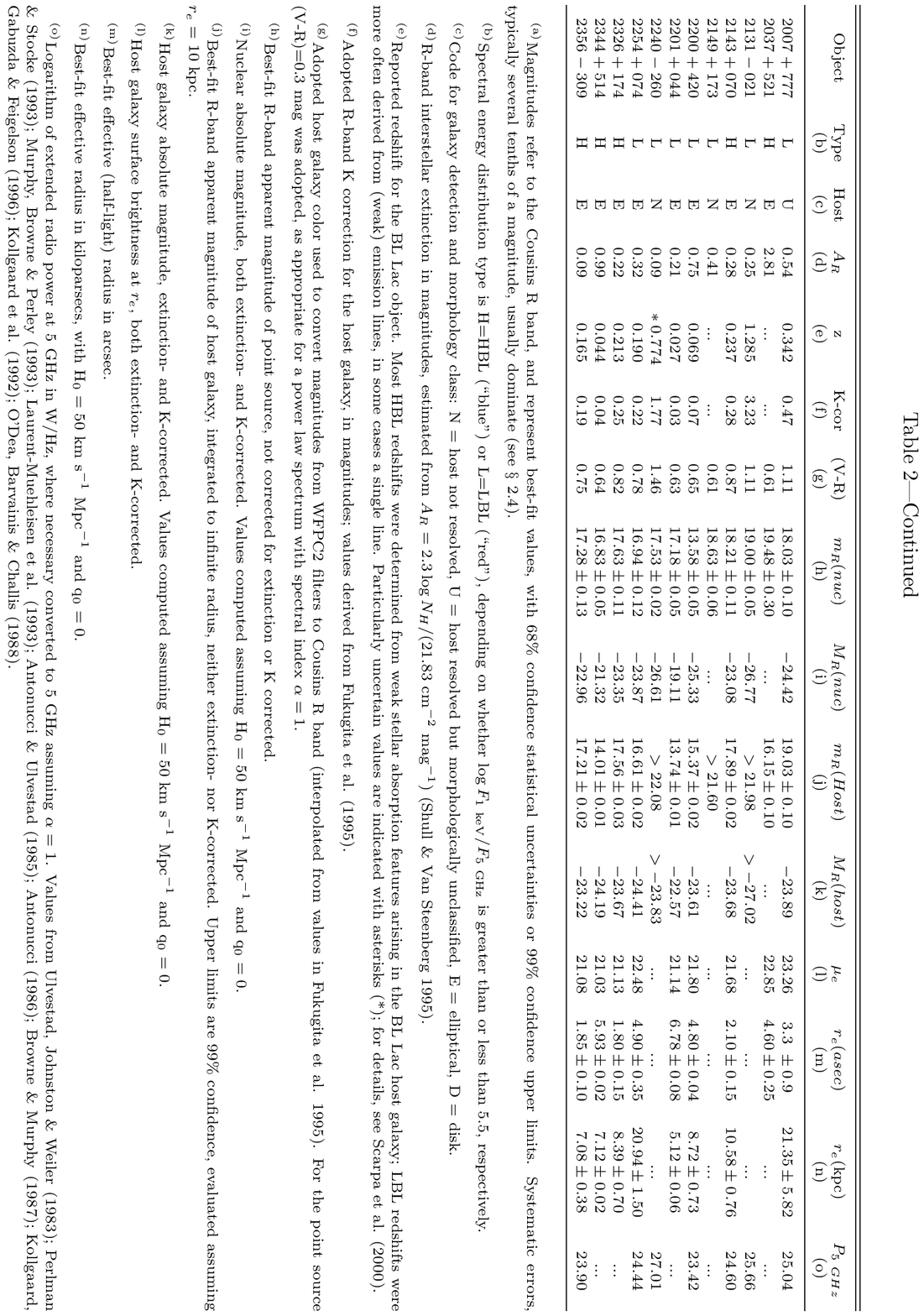}
  }
\end{figure}

\end{document}